\documentclass[11pt]{article}
\usepackage{moriond,epsfig}

\def\be{\begin{equation}}
\def\ee{\end{equation}}
\def\bea{\begin{eqnarray}}
\def\eea{\end{eqnarray}}

\begin{document}
\vspace*{4cm}
\title{RECENT STRUCTURE FUNCTION RESULTS FROM CCFR}

\author{ B.~T.~Fleming,$^{2}$ T.~Adams, $^{4}$ A.~Alton,$^{4}$
C.~G.~Arroyo,$^{2}$ S.~Avvakumov,$^{7}$ L.~de~Barbaro,$^{5}$
P.~de~Barbaro, $^{7}$ A.~O.~Bazarko,$^{2}$ R.~H.~Bernstein,$^{3}$
A.~Bodek,$^{7}$ T.~Bolton, $^{4}$ J.~Brau,$^{6}$ D.~Buchholz,$^{5}$
H.~Budd,$^{7}$ L.~Bugel,$^{3}$ J.~Conrad,$^{2}$ R.~B.~Drucker,$^{6}$
J.~A.~Formaggio,$^{2}$ R.~Frey,$^{6}$ J.~Goldman,$^{4}$
M.~Goncharov,$^{4}$ D.~A.~Harris,$ ^{7} $ R.~A.~Johnson,$^{1}$
J.~H.~Kim,$^{2}$ B.~J.~King,$^{2}$ T.~Kinnel,$ ^{8}$
S.~Koutsoliotas,$^{2}$ M.~J.~Lamm,$^{3}$ W.~Marsh,$^{3}$
D.~Mason,$^{6}$ K.~S.~McFarland, $^{7}$ C.~McNulty,$^{2}$
S.~R.~Mishra,$^{2}$ D.~Naples,$^{4}$ P.~Nienaber,$^{3}$
A.~Romosan,$^{2}$ W.~K.~Sakumoto,$^{7}$ H. Schellman,$^{5}$
F.~J.~Sciulli,$^{2}$ W.~G.~Seligman,$^{2}$ M.~H.~Shaevitz,$^{2}$
W.~H.~Smith,$^{8}$ P.~Spentzouris,$^{2}$ E.~G.~Stern,$^{2}$
M.~Vakili,$^{1}$ A.~Vaitaitis,$^{2}$ M.~Vakili,$^{1}$ U.~K.~Yang,$^{7}$
J.~Yu,$^{3}$ G.~P.~Zeller,$^{5}$ and E.~D.~Zimmerman$^{2}$\\ (The
CCFR/NuTeV Collaboration)}

\address{
$^{1}$ Univ. of Cincinnati, Cincinnati, OH 45221;
$^{2}$ Columbia University, New York, NY 10027 
$^{3}$ Fermilab, Batavia, IL 60510
$^{4}$ Kansas State University, Manhattan, KS 66506 
$^{5}$ Northwestern University, Evanston, IL 60208;
$^{6}$ Univ. of Oregon, Eugene, OR 97403 
$^{7}$ Univ. of Rochester, Rochester, NY 14627;
$^{8}$ Univ. of Wisconsin, Madison, WI 53706}

\maketitle \abstracts{A new structure function analysis of CCFR deep inelastic $\nu$-N and
$\overline{\nu}$-N scattering data is presented for previously
unexplored kinematic regions down to Bjorken $x=0.0045$ and
$Q^{2}=0.3$ GeV$^{2}$.  Comparisons to charged lepton scattering data
from NMC \cite{NMC} and E665\cite{E665} experiments are made and the
behavior of the structure function $F_{2}^{\nu}$ is studied in the
limit $Q^{2} \rightarrow 0$.}

Neutrino structure function measurements in the low Bjorken $x$, low
$Q^2$ region can be used to study the axial-vector component of the
weak interaction as well as to test the limits of parton distribution
universality.  We present a first measurement of the
structure function $F_2$ in neutrino scattering, from the CCFR data,
for $Q^2<1$ GeV$^{2}$ and $0.0045 < x < 0.035 $.  In this region where
perturbative and non-perturbative QCD meet, we present a
parameterization of the data which allows us to test the partially
conserved axial current (PCAC) limit of $F_2$ in neutrino scattering.

The universality of parton distributions can be tested by comparing
neutrino scattering data to charged lepton scattering data.  Past
measurements for $0.0075<x<0.1$ and $Q^{2}>1.0$ GeV$^{2}$ have
indicated that $F_{2}^{\nu }$ differs from $F_{2}^{\mu }$ by 10-15\%
\cite{seligman:1997mc}.  This discrepancy has been partially resolved
by recent analyses of $F_{2}^{\nu}$ at $Q^2 > 1.0 $
GeV$^{2}$ \cite{unki,thomas}.  While we expect and have now observed that
parton distribution universality holds in this region, this need not
be the case at lower values of $Q^{2}$.  Deviations from this
universality at lower $Q^{2}$ are expected due to differences in
vector and axial components of electromagnetic and weak interactions.
In particular, the electromagnetic interaction has only a vector
component while the weak interaction has both vector and axial-vector
components.  Vector currents are conserved (CVC) but axial-vector
currents are only partially conserved (PCAC).  Adler \cite{adler}
proposed a test of the PCAC hypothesis using high energy neutrino
interactions, a consequence of which is the prediction that $F_{2}$
approaches a non-zero constant as $Q^{2} \rightarrow 0$ due to U(1)
gauge invariance.  A determination of this constant is performed here
by fitting the low $Q^{2}$ data to a phenomenological curve developed
by Donnachie and Landshoff \cite{DL}.

In previous analyses a slow rescaling correction was applied to
account for massive charm effects.  This is not applied here since the
corrections are model dependent and uncertain in this kinematic range.
As a result, neutrino and charged lepton DIS data must be compared
within the framework of charm production models, accomplished by
plotting the ratio of data to theoretical model.  The theoretical
calculation corresponding to the CCFR data employs NLO QCD including
heavy flavor effects as implemented in the TR-VFS(MRST99) scheme
\cite{mfs,mrst}.  The theoretical calculation corresponding to NMC and
E665 data is determined using TR-VFS(MRST99) for charged lepton
scattering.  Other theoretical predictions such as ACOT-VFS(CTEQ4HQ)
\cite{vfs,acot} and FFS(GRV94) \cite{ffs} do not significantly change
the comparison.  For acceptance, smearing, and radiative corrections
we chose an appropriate model for the low $x$, low $Q^{2}$ region, the
GRV \cite{GRV} model of the parton distribution functions.  The GRV
model is used up to $Q^{2}=1.35$ GeV$^{2}$ where it is normalized to a
LO parameterization \cite{BG} used above this.  Finally, a correction
is applied for the difference between $xF_{3}^{\nu}$ and
$xF_{3}^{\overline{\nu}}$, determined using a LO calculation of
$\Delta xF_{3}= xF_{3}^{\nu} - xF_{3}^ { \overline{\nu}}$.  The recent
CCFR $\Delta xF_{3}$ measurement \cite{unki} is higher than this LO
model \cite{BG} and all other recent LO and NLO theoretical
predictions in this kinematic region.  An appropriate systematic error
is applied to account for the differences between the theory and this
measurement.

\begin{figure}[t]
\hfill
\begin{minipage}{5.5 cm}
\psfig{figure=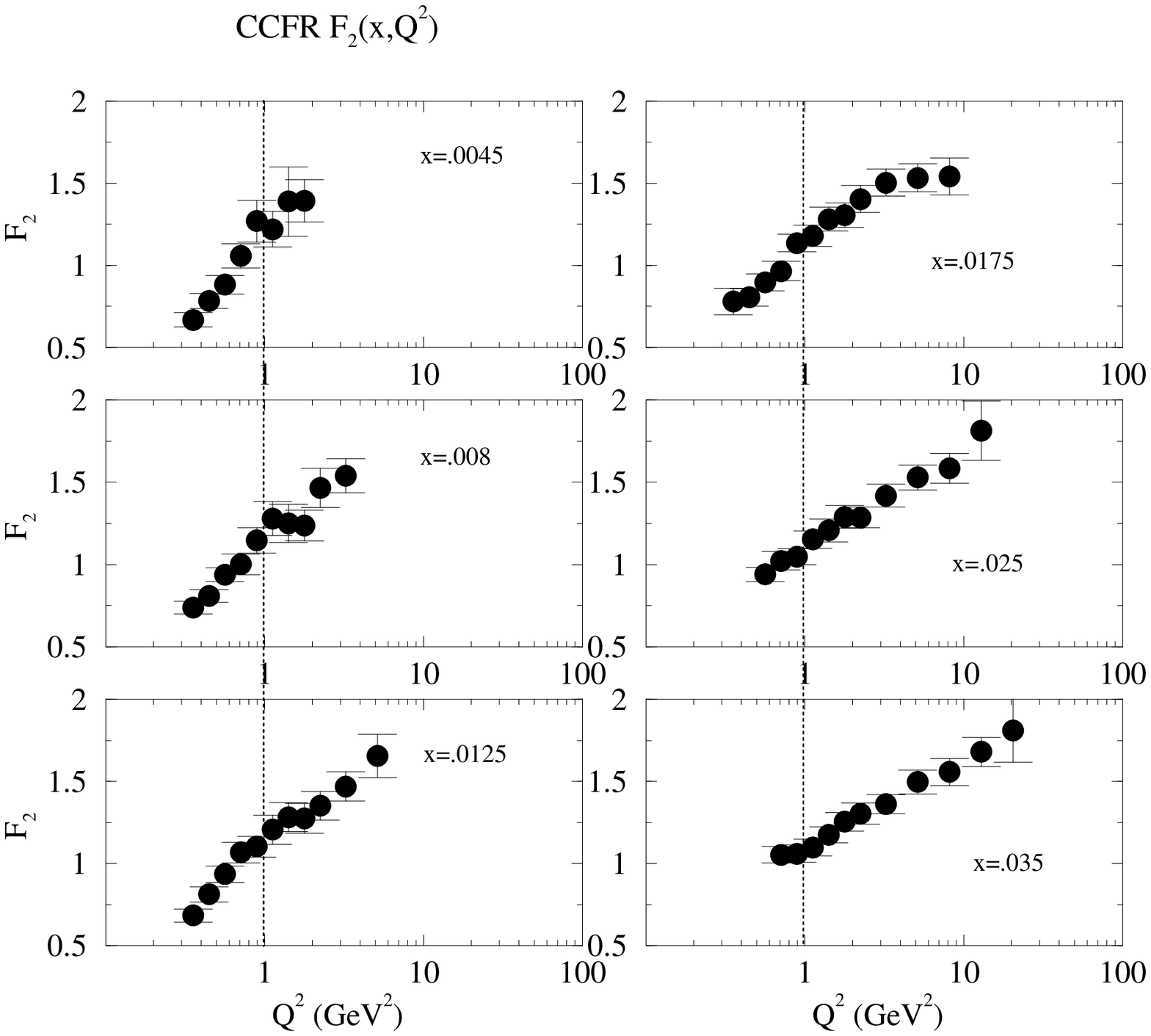,width=2.8in}
\caption{CCFR $F_{2}$ at low $x$, low $Q^{2}$.  Data to the left of the vertical line at $Q^{2}=1.0$ represent the new kinematic regime for this analysis.}
\label{fig:SF}
\end{minipage}
\hfill
\begin{minipage}{5.5cm}
\psfig{figure=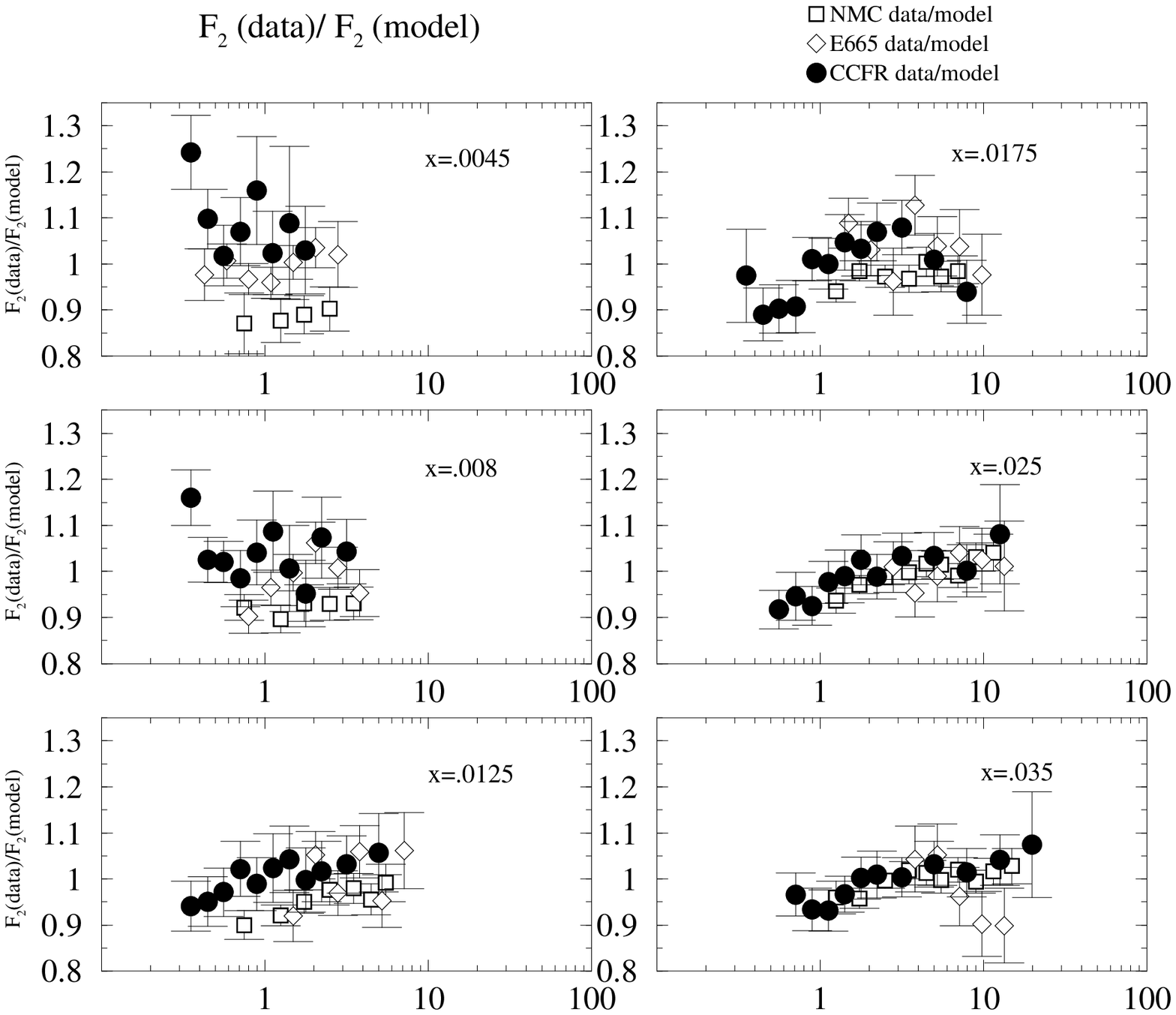,width=2.8in}
\caption{$F_{2}$ data/theory from CCFR $\nu $-Fe DIS compared to
$F_{2}$ from NMC and E665 DIS. Errors bars are statistical and
systematic added in quadrature. Theoretical predictions are those of
TR-VFS(MRST99).}
\label{fig:compare}
\end{minipage}
\hfill\mbox{}
\end{figure}

The combination of the inclusion of the GRV model at low $x$ and low
$Q^{2}$, its effect on the radiative corrections, and removal of the
slow rescaling correction help to resolve the longstanding discrepancy
between the neutrino and charged lepton DIS data
above $x=0.015$. $F_{2}$ is plotted in Figure~\ref{fig:SF}.  Errors
are statistical and systematic added in quadrature. A line is drawn at
$Q^{2}=1$ GeV$^{2}$ to highlight the kinematic region this analysis
accesses.  Figure \ref{fig:compare} compares $F_{2}$ (data/theoretical
model) for CCFR, NMC, and E665.  There is agreement to within 5\% down
to $x=0.0125$.  Below this, as $x$ decreases, CCFR $F_{2}$
(data/theory) becomes systematically higher than NMC $F_{2}$
(data/theory). Differences between scattering via the weak interaction
and via the electromagnetic interaction as $Q^{2} \rightarrow 0$ may
account for the disagreement in this region.

\begin{figure}[h]
\hfill
\begin{minipage}{5.5cm}
\epsfig{figure=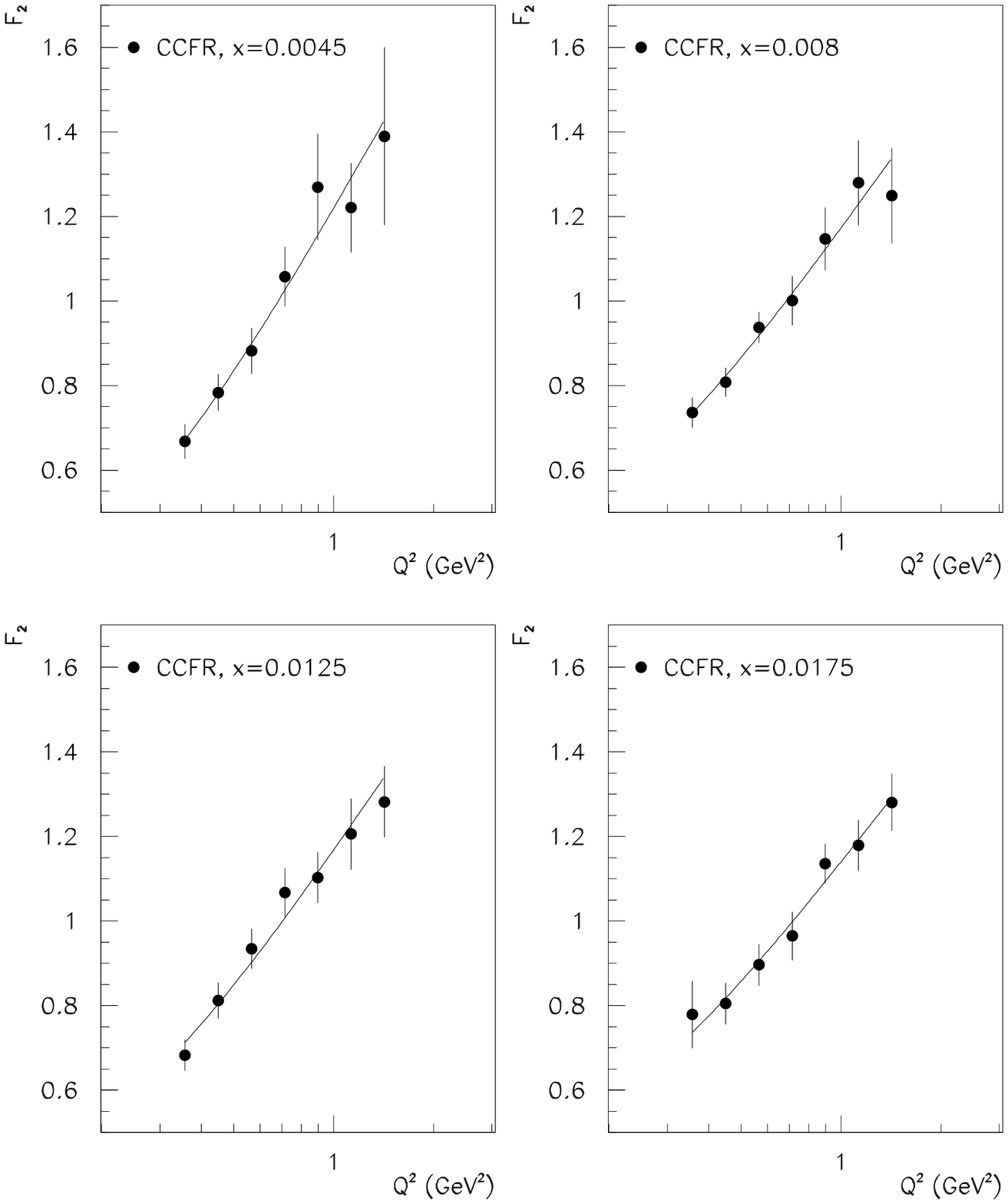,width=3.0in}
\caption{Results from fit to CCFR data to extrapolate to $F_{2}(Q^{2}=0)$.}
\label{fig:fits}
\end{minipage}
\hfill
\begin{minipage}{5.5cm}
\epsfig{figure=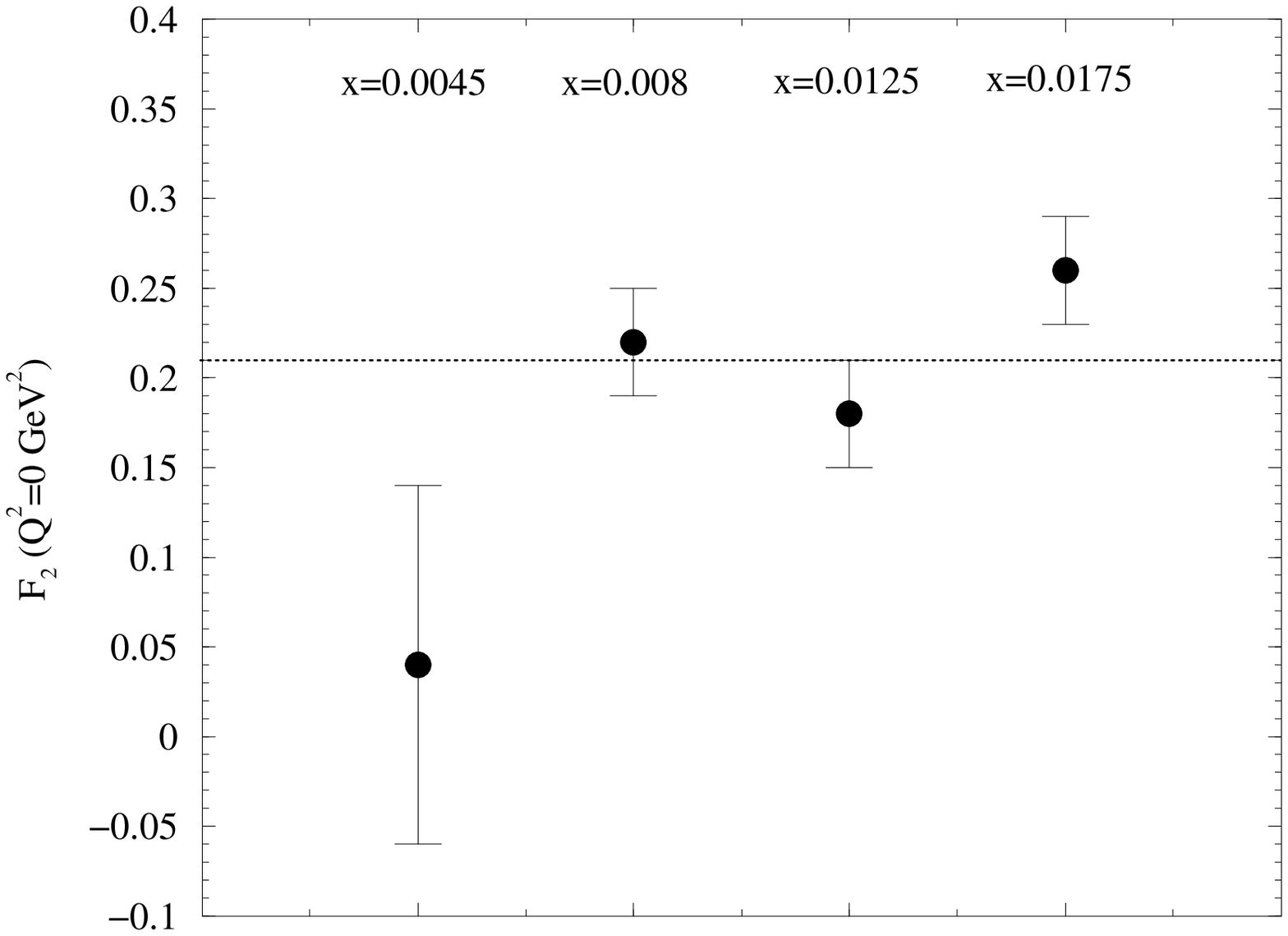,width=3.0in}
\caption{$F_{2}(Q^{2}=0$ GeV$^{2})$ from different $x$ bins.  A line is drawn at the weighted average of all four measurements.}
\label{fig:fitsplot}
\end{minipage}
\hfill\mbox{}
\end{figure}

In charged lepton DIS, the structure function
$F_{2}$ is constrained by gauge invariance to vanish with
$Q^{2}$ as $Q^{2}\rightarrow 0$.  Donnachie and Landshoff predict that in the
low $Q^{2}$ region, $F_{2}^{\mu}$ will follow the form \cite{DL}:
\begin{equation}
C \left( \frac{Q^{2}}{Q^{2}+A^{2}} \right) .
\label{eq:nmc_e665}
\end{equation}
However, in the case of neutrino DIS, the axial
component of the weak interaction may contribute a nonzero component to
$F_{2}$ as $Q^{2}$ approaches zero.  Donnachie and Landshoff predict
that $F_{2}^{\nu}$ should follow a form with a non-zero contribution
at $Q^{2}=0$:
\begin{equation}
\frac{C}{2} \left( \frac{Q^{2}}{Q^{2}+A^{2}} + \frac{Q^{2} + D}{Q^{2}
+ B^{2}} \right) .
\end{equation}  
Using NMC and E665 data, corrected in this case to be equivalent to
scattering from an iron target using a parameterization of SLAC Fe/D
data \cite{WGSthesis}, we do a combined fit to the form predicted for
$\mu$ DIS and extract the parameter $A = 0.81
\pm 0.02$ with $\chi ^{2}/DOF = 27/17$.  The error on $A$ is
incorporated in the systematic error on the final fit.  Inserting this
value for $A$ into the form predicted for $\nu$N DIS, we fit CCFR data to extract parameters B, C, and D, and
determine the value of $F_{2}$ at $Q^{2}=0$.  Only data below
$Q^{2}=1.4$ GeV$^{2}$ are used in the fits.  The CCFR $x$-bins that
contain enough data to produce a good fit in this $Q^{2}$ region are
$x=0.0045$, $x=0.0080$, $x=0.0125$, and $x=0.0175$.  Figure
\ref{fig:fits} and Table \ref{tab:fitresults} show the results of the
fits.  Error bars consist of statistical and systematic terms added in
quadrature but exclude an overall correlated normalization uncertainty
of 1-2\%. The values of $F_{2}$ at $Q^{2}=0$ GeV$^{2}$ in the three
highest $x$-bins are statistically significant and are within $1
\sigma$ of each other.  The lowest $x$ bin has large error bars but is
within $1.5 \sigma$ of the others.  Taking a weighted average of the
parameters $B, C, D,$ and $F_{2}$ yields $B=1.53 \pm 0.02, C=2.31 \pm
0.03, D=0.48 \pm 0.03$, and $F_{2}(Q^{2}=0)=0.21 \pm 0.02$.  Figure
\ref{fig:fitsplot} shows $F_{2}(Q^{2}=0)$ for the different $x$ bins.
Inclusion of an $x$ dependence of the form $x^{\beta}$ does not change
the overall fits or $\chi ^{2}$s.  However, the Donnachie and
Landshoff mass parameter, $B$, appears to depend on $x$, with higher
values corresponding to higher $x$.  Thus, $F_{2}$ at higher $x$
approaches $F_{2}(Q^{2}=0)$ more slowly than at lower $x$.

In summary, a comparison of $F_{2}$ from neutrino DIS to that from
muon DIS shows good agreement above $x=0.0125$, but shows differences
at smaller $x$. This low $x$ discrepancy can be explained by the
different behavior of $F_{2}$ from $\nu$ DIS to that from $e/\mu$ DIS
as $Q^{2} \rightarrow 0$.  CCFR $F_{2}^{\nu}$ data favors a non-zero
value for $F_{2}$ as $Q^{2} \rightarrow 0$.

We would like to thank Fred Olness for many useful discussions.  \cite{fred}

\begin{table}[h]
\caption{Fit results for CCFR data. CCFR data is fit to Eq. 4 with $A =
0.81 \pm 0.02$ as determined by fits to NMC and E665 data.  B, C, D, and
$F_{2}$ at $ Q^{2}=0$ results shown below. $N=4$ for all fits.}
\begin{center}
\begin{tabular}{ccccccc}
$x$ & $B$ & $C$ & $D$ & $F_{2}^{\nu}(Q^{2}=0)$ & $\chi^{2}/N$\\
\hline \hline $0.0045$ & $1.49 \pm 0.02$ & $2.62 \pm 0.26$ & $0.06
\pm 0.17$ & $0.04 \pm 0.10$ & $0.5$\\ 
$0.0080$ & $1.63 \pm 0.05$ & $2.32 \pm 0.05$ & $0.50 \pm 0.05$ & $0.22 \pm 0.03$ & $0.5$\\
$0.0125$ & $1.63 \pm 0.05$ & $2.39 \pm 0.05$ & $0.40 \pm 0.05$ & $0.18
\pm 0.03$ & $1.0$\\ $0.0175$ & $1.67 \pm 0.05$ & $2.20 \pm 0.05$ &
$0.65 \pm 0.07$ & $0.26 \pm 0.03$ & $0.5$\\
\end{tabular} 
\end{center}
\label{tab:fitresults}
\end{table}



\end{document}